# Adsorption of Water at the SrO Surface of Ruthenates

Daniel Halwidl[1+], Bernhard Stöger[1+], Wernfried Mayr-Schmölzer[1,2], Jiri Pavelec[1], David Fobes[3], Jin Peng[3], Zhiqiang Mao[3], Gareth Parkinson[1], Michael Schmid[1], Florian Mittendorfer[1,2], Josef Redinger[1,2], Ulrike Diebold[1*]

[1]Institute of Applied Physics, TU Wien, Wiedner Hauptstrasse 8-10/134, A-1040 Vienna, Austria

[2]Center for Computational Materials Science, TU Wien, Wiedner Hauptstrasse 8-10/134, A-1040 Vienna, Austria

[3]Department of Physics and Engineering Physics, Tulane University, New Orleans, LA 70118, USA

Perovskite oxides hold high promise in applications ranging from solid oxide fuel cells to catalysts, but their surface chemistry is poorly understood. Here the interaction of water with the (001) surfaces of $Sr_{n+1}Ru_nO_{3n+1}$ (n = 1, 2) is investigated with low-temperature Scanning Tunneling Microscopy (STM), X-ray Photoelectron Spectroscopy (XPS), and Density Functional Theory (DFT) based calculations. These layered perovskites cleave between neighboring SrO planes, yielding almost ideal, rocksalt-like surfaces. An adsorbed water monomer dissociates and forms a pair of hydroxide ions $(OH)_{ads}$ + $O_{surf}H$. The $(OH)_{ads}$ stays trapped at Sr-Sr bridge positions, circling the $O_{surf}H$ with a measured activation energy of 187±10 meV. For higher coverage the simple rocksalt analogy breaks down because neighboring adsorption sites are rendered symmetrically inequivalent by the rotation of the $RuO_6$ octahedra in the second layer. Water dimers assemble into one-dimensional chains upon gentle annealing, forming a percolating network at high coverage. Molecular water adsorbs in the gaps, resulting in a mixed mode adsorption.

---

[+] These authors contributed equally to the work
[*] Corresponding author: diebold@iap.tuwien.ac.at

Perovskite oxides, ternary compounds with the principal formula $ABO_3$, show a large variation in their composition and structure, which leads to an almost unlimited flexibility of their physical and chemical properties. In particular perovskite-type materials that are categorized as mixed ionic and electronic conductors (MIECs) are useful in a wide variety of energy-conversion devices. They serve as the air electrode in solid oxide fuel cells (SOFC) [1-4] and solid oxide electrolysis cells (SOEC) [5], where they are also increasingly discussed as the fuel electrode [6,7]; as ion separation membranes in carbon capturing schemes [8]; and as catalysts in solar and thermochemical $H_2$ and CO production and in air batteries [9,10]. To enable a rational design of better materials [10,11] one needs to understand the underlying surface chemistry. Compared to the comparatively simpler systems used in heterogeneous [12,13] and low-temperature electrocatalyis [14], knowledge of the gas-surface interaction at the molecular level is seriously underdeveloped for these complex materials.

The reactivity of any solid depends on the structure and composition of its top atomic layers. There is overwhelming evidence that many perovskites are mostly A-cation terminated under operating conditions used for electrochemical energy conversion [15-19]. Thus an AO-terminated perovskite is a natural place to start investigating the fundamentals of perovskite surface chemistry; here we use the SrO surface that results from cleaving layered strontium ruthenate crystals as an ideal model system.

We focus on the interaction of this surface with $H_2O$. The adsorption configuration of this molecule is essential in high[7] or low[10] temperature water electrolysis and in thermochemical water splitting[9]. It is ubiquitous in the environment [20-22] which has serious consequences for the degradation of SOFC electrodes [23-25]. For the comparatively simple binary AO oxides, many of the details have been worked out both theoretically and experimentally [26-33]. This raises an intriguing question: can the concepts derived for rocksalt oxides also be applied to – in principle much more complex – perovskite surfaces? To answer this question we follow the formation of a water layer from the single-molecule limit to the full monolayer. We find that an isolated water molecule on our SrO terminated surface behaves exactly as expected: $H_2O$ dissociates, and with STM we observe an intriguing dynamic behaviour that has been predicted in ab-initio molecular dynamics calculations [26]. The interaction between neighbouring molecules, however, is affected by rotation and tilting of the octahedra surrounding the Ru atoms in the second layer. This influences both the short-range and long-range ordering that evolves with coverage, and helps explain why water adsorbs as an intact molecule as the overlayer fills in. Our detailed STM and DFT studies provide a clear interpretation of the complicated XPS spectra of this material, and a benchmark for investigations of more complex materials at higher pressures.

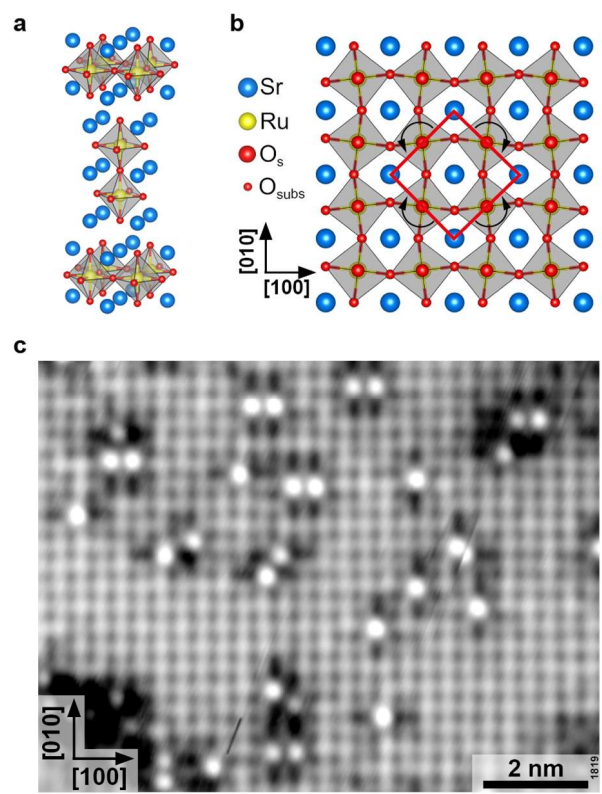

**Figure 1. Water adsorption at cleaved strontium ruthenate single crystals.** (a) Unit cell of the $n = 2$ member of the $Sr_{2n+1}Ru_nO_{3n+1}$ Ruddlesden-Popper series. The crystal easily cleaves between neighboring SrO layers. (b) Top view of the $Sr_3Ru_2O_7(001)$ surface. While the surface termination is similar to rocksalt SrO, the rotation of the underlying $RuO_2$ octahedra gives rise to a so-called c(2×2) structure. c) Scanning tunneling microscopy image ($T_{sample}$ = 78 K, $V_{sample}$ = +0.05 V, $I_{tunnel}$ = 0.15 nA, image rotated 69° counterclockwise to scan direction) of 0.05 Langmuir of water, dosed at 115 K. Water monomers appear as bright, isolated spots. Some adsorb next to each other as 'dimers'.

**Water monomer adsorption and dynamics**

As samples we used strontium ruthenate single crystals $Sr_{n+1}Ru_nO_{3n+1}$ ($n$ = 1, 2) that are part of the Ruddlesden-Popper series. These consist of $n$ perovskite-like $SrRuO_3$ layers, separated by two layers of SrO (Fig. 1a) that easily cleave apart [34]. While the resulting surfaces resemble the (001) facets of rocksalt SrO, there are also important differences. With Sr-Sr separations of 3.9 Å, the lattice constant is expanded as compared to the 3.6 Å in SrO. The octahedral units containing the Ru atoms are rotated alternatingly by 8.5±2.5°, which gives rise to an apparent c(2×2) structure for $Sr_2RuO_4$ [35]; a similar octahedral rotation (8.1°) is inherent in the bulk $Sr_3Ru_2O_7$ lattice [36], see Fig. 1b.

We start by analyzing a very low coverage of water on $Sr_2RuO_4$ (001). Figure 1c shows the surface after exposure of 0.05 Langmuir (L, where 1 L equals an exposure to 1 x $10^{-8}$ mbar for 133 s) of water at 115 K. At these low temperatures, migration of the molecules is largely, albeit not completely, suppressed. Monomers are visible that are separated by a distance that is a large enough to safely consider them isolated, but a few 'dimers' are already observed. In such atomically-resolved STM images, the surface Sr and O atoms are imaged as bright protrusions and dark small dots, respectively [34]. Each monomer is situated between two Sr atoms. The presence of the monomer affects the electronic structure of the surroundings, which is manifested in a changed contrast in the STM image, as has been seen for other adsorbates on this surface [34,37]. The change is not symmetric; the darker patches are more pronounced on one side.

A single water molecule should dissociate on this surface according to trends for rocksalt oxides that have been worked out in DFT calculations [26,28]. For MgO molecularly adsorbed water is predicted to be more stable, adsorbing in an on-top configuration on an Mg cation site. With increasing lattice constant, dissociative adsorption should occur, which results in a pair of hydroxyl ions $(OH)_{ads}$ + $O_{surf}H$. This is facilitated by the increase in structural flexibility in the series CaO, SrO, and BaO [28]. Indeed, in our own DFT calculations of 1/16 monolayers (ML) water on the $Sr_3Ru_2O_7$ surface, the adsorption energy of the dissociated molecule ($E_{ads}$ = 1.26 eV) is higher than for the intact molecule ($E_{ads}$ = 1.08 eV).

When testing these theoretical predictions experimentally, one runs into the difficulty that such investigations need to be conducted under conditions where the monomer stays isolated and is not affected by water-water interaction that can also lead to dissociation [29,32]. This precludes many spectroscopic methods for intensity reasons. For MgO the prediction of molecularly adsorbed water has been verified experimentally using the tip of an STM as a tool for selective dissociation [30]. Here the expected dissociation is ascertained through the observation of 'dynamic ion pairs' that have been predicted theoretically in earlier calculations [26,27,38].

The geometry of the dissociated water is shown in Fig. 2a. The $(OH)_{ads}$ fragment is adsorbed in a Sr-Sr bridge site with an Sr-O distance of 2.60 Å, while the remaining H is transferred to a neighboring surface oxygen atom $O_{surf}$. The OH bond length of 0.97 Å for the $(OH)_{ads}$ fragment is slightly smaller than the value of 1.03 Å found for the $O_{surf}$H bond. The latter bond is tilted to allow for the formation of an additional hydrogen bond to the $(OH)_{ads}$ fragment, with an H-$(OH)_{ads}$ distance of 1.53 Å. According to Tersoff-Haman simulations the $(OH)_{ads}$ gives rise to a bright feature surrounded by a dark halo in the STM measurements (Supplementary Fig. 2).

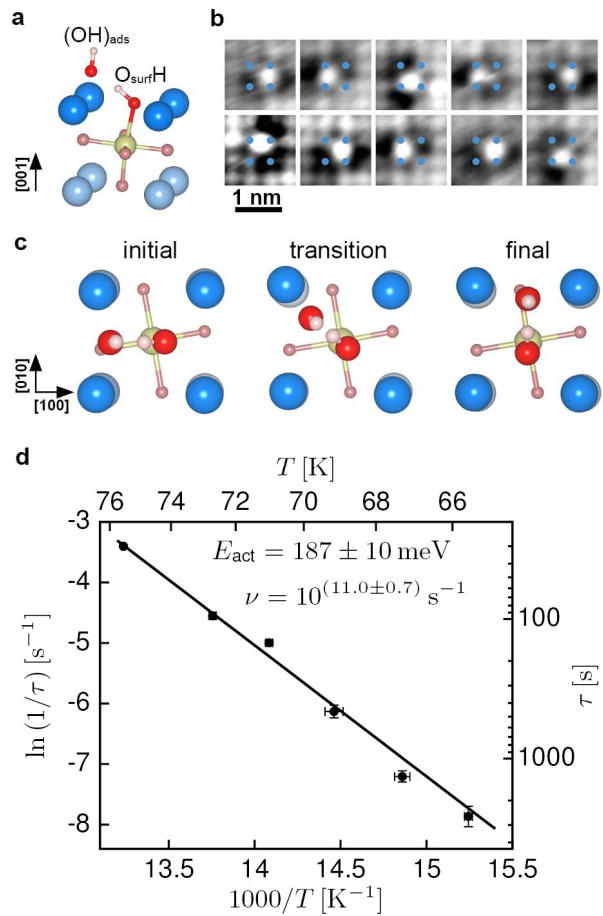

**Figure 2. Dissociated water forming a 'dynamic ion pair'.** a) Adsorption geometry of the lowest-energy configuration of a water monomer on the SrO-terminated surface of $Sr_3Ru_2O_7$. The $(OH)_{ads}$ and the $O_{surf}H$ interact through a H bond, preventing separation of the two fragments. b) A few selected, consecutive images for water monomer motion at 78 K; the $(OH)_{ads}$ hops between equivalent bridge sites, circling the $O_{surf}H$. c) The hopping modeled with DFT, showing the initial state, transition state, and final state; the calculations yield an activation energy of 171 meV. d) Arrhenius plot for this motion obtained from time-lapse STM movies shown in the Supplement. An activation energy of $E_{act} = 187 \pm 10$ meV and a prefactor $\nu = 10^{(11.0\pm0.7)}$ s$^{-1}$ is derived, see SI.

Consecutive images of the same sample area show that the $(OH)_{ads}$ is not immobile, but hops from one Sr-Sr bridge site to the next, orbiting the $O_{surf}H$ group (see Fig 2b, and movies M1-M6 in the Supplement). The DFT results (Fig. 2c) show how this hopping occurs. The $(OH)_{ads}$ is captured by the hydrogen bond formed by the $O_{surf}H$, which results in a rotation around this group.

Consequently the orientation of the $O_{surf}H$ group follows the movement of the $(OH)_{ads}$ during a hop to the neighboring bridge site. The DFT calculations predict an activation energy, $E_{act}$, of 171 meV for this process. Series of STM images were taken at various temperatures and evaluated using a simple model for one-dimensional diffusion [39]. The average time between two hops was measured (see Fig. 2d and the Supplement). Fitting the Arrhenius equation gives an attempt frequency of $10^{(11.0\pm0.7)}$ s$^{-1}$ and an $E_{act}$ of 187±10 meV, in good agreement with the DFT calculations.

**Water dimers and chains**

In addition to monomers, some 'dimers' are observed already at low temperatures (see Fig. 1c), i.e., two bright spots located at neighboring Sr-Sr bridge sites. This is consistent with DFT results that show an energy gain of about 109 meV/molecule for dissociative adsorption next to each other (1/8 ML) compared to more separated sites (1/16 ML). No hopping is observed, apparently the circular motion is hindered by the second, neighboring adsorbate. Interestingly, all dimers are situated in one specific position of the c(2×2) surface, see the grid overlay in Supplementary Figs. 4 and 5. The tendency for water molecules to adsorb at specific, neighboring sites is more pronounced when a small coverage of water is annealed. When the temperature is raised, the water molecules become mobile and aggregate into one-dimensional chains, see Fig. 3. On the $Sr_3Ru_2O_7$ surface the chains consist of an assembly of dimers. They appear slightly displaced from the Sr-Sr bridge site. The $(OH)_{ads}$ within each dimer tilt towards the same side, suggesting that the associated $O_{surf}H$ sit next to each other. In Fig. 3a horizontally oriented chains with an upwards/downwards shifted position are marked by a blue/red box. The shift is discerned most easily when two differently tilted dimers sit next to each other (Fig. 3a, mixed-color box and inset). The strict alignment in the chains is pronounced on the $Sr_3Ru_2O_7$ surface, where the dimerization is dictated by the octahedral rotation inherent in the bulk structure. For the one-layer perovskite, $Sr_2RuO_4$, water also assembles in chains, but the structure is less rigid, and a zig-zag configuration is also observed (Fig. 3b).

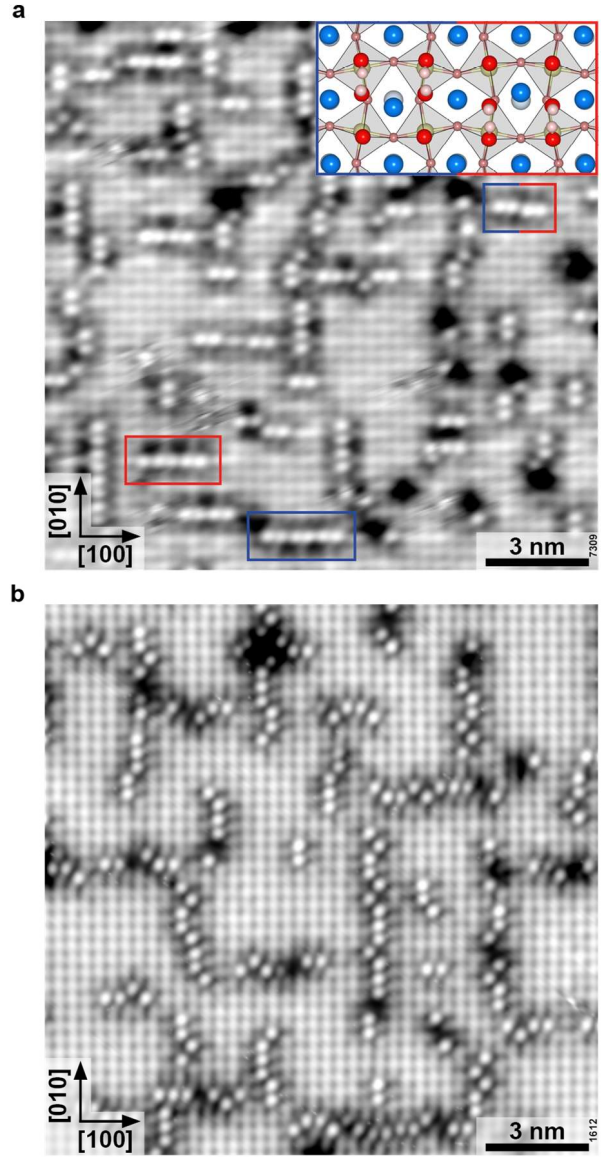

**Figure 3. Formation of H₂O chains.** STM images after annealing a H$_2$O dose of 0.05 L to room temperature for 1 hour. a) Sr$_3$Ru$_2$O$_7$ ($T$ = 78 K, $V_{\text{sample}}$ = +0.5 V, $I_{\text{tunnel}}$ = 0.15 nA, image rotated 20° counterclockwise to scan direction). Dimers within the chains are slightly shifted upwards (blue rectangle) and downwards (red rectangle) from the Sr-Sr bridge sites. The inset shows the adsorption geometry of two dimers sitting next to each other, where the O$_{\text{surf}}$H fragments of the left (right) dimer sit above (below) a row of Sr atoms. b) Sr$_2$RuO$_4$ ($T$ = 78 K, $V_{\text{sample}}$ = +0.7 V, $I_{\text{tunnel}}$ = 0.15 nA, image rotated 36.5° clockwise to scan direction).

The tendency to adsorb in well-defined positions in the c(2x2) structure is related to the symmetry of the underlying perovskite lattice. Due to the clockwise and counterclockwise rotation of the edge-sharing $RuO_6$ octahedra, the 'empty space' between them resembles rhombs with their long diagonals rotated alternatingly by 90° in a top view model (Fig 1b and inset Fig. 3a). According to DFT calculations, two neighboring, dissociated water molecules are energetically more favorable (by $\Delta E = 62$ meV) at positions across the short diagonal of these rhombs (see Supplementary Figure 3). The energy difference results from the interaction between the Sr atom and the two $(OH)_{ads}$. When the dimers span the "short" site, the Sr atom between two $(OH)_{ads}$ can relax away as it moves with the $RuO_6$ octahedra rotation direction, while the Sr would move against the rotation of the octahedra at the other position. The preference of the dimers for a specific site of the c(2×2) grid is the same for $Sr_3Ru_3O_7$ and $Sr_2RuO_4$. This can be explained by the similar surface structure: For $Sr_2RuO_4$, the octahedra are not rotated in the bulk. However, due to a surface reconstruction [35] the surface ocatahedra are rotated by 8.5±2.5°, which is quite close to the octahedral rotation inherent in the $Sr_3Ru_2O_7$ lattice (8.1°).

**Full water monolayer**

Figure 4a shows the structure after the $Sr_2RuO_4$ sample was exposed to 1 L of water at 160 K. The chains now form a percolating network. (A water overlayer on $Sr_3Ru_2O_7$ has a similar appearance, see Supplementary Figure 6.) In the gaps between the bright lines of the chains, water appears in the form of isolated, bright spots. Here the water appears in the form of isolated, bright spots; their number (up to six) depends on the size of the gap. The small-area image in Fig. 4b shows the spots in relationship to the underlying surface atoms. Under the assumption that the $(OH)_{ads}$ in the chains continue to be positioned in Sr-Sr bridge positions, the bright spots are located mostly on-top of Sr.

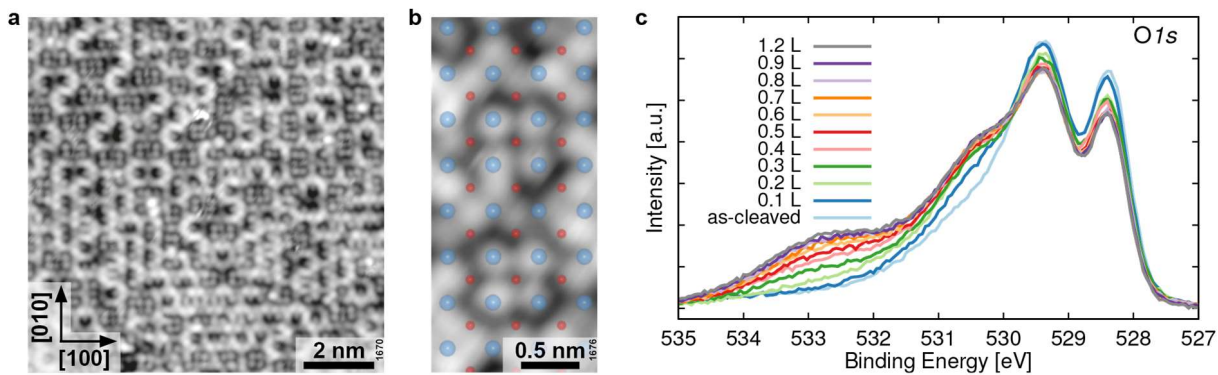

**Figure 4. Full monolayer of water.** a) STM image ($T$ = 78 K, $V_{sample}$ = +0.5 V, $I_{tunnel}$ = 0.1 nA, image rotated 53.5° counterclockwise to scan direction) of $Sr_2RuO_4$ after dosing 1.0 L water at 160 K. b) Small-area STM image ($T$ = 78 K, $V_{sample}$ = +0.4 V, $I_{tunnel}$ = 0.1 nA) with substrate atoms (blue – Sr, red – O) overlaid. Note the switch in adsorption site, from Sr-Sr positions within the 1D lines, to on-top configuration in the gaps in between. c) XPS spectra of the O*1s* region after exposure of the clean, cleaved $Sr_2RuO_4$ surface to increasing amounts of water at 140 K. The shoulder and peak at 530.4 eV and 532.8 eV are indicative of hydroxyls and molecular water, respectively.

The calculations in ref. [28] suggest that a switch in adsorption site from bridge to top position indicates molecular instead of dissociative adsorption. This was tested with x-ray photoelectron spectroscopy (XPS, Fig. 4c). The O*1s* region of the clean, as-cleaved $Sr_2RuO_4$ surface shows a two-peak structure. In ref. [40] it was proposed that this should be due to the different bonding environments of the apical and the in-plane O atoms within the $RuO_6$ octahedra; this is consistent with our calculations, see Supplementary Table 4. Upon dosing a small amount of $H_2O$ a shoulder at 530.4 eV appears that is indicative of OH groups. The substrate O1*s* peak at 528.4 eV decreases more strongly, in agreement with the assignment of apical O atoms, i.e., the ones residing within the top surface plane. With increasing exposure a clearly distinct peak at 532.8 eV develops, consistent with our theoretical predictions for the presence of molecular water, see Supplementary Table 4. Additional water leads to features indicative of water multilayers in XPS, see Supplementary Fig. 8. Annealing to room temperature is sufficient to remove most of the adsorbed water except for a network of water chains that remains on the surface (Supplementary Fig. 7).

**Adsorption at oxygen vacancies**

The surface chemistry of oxides is often heavily influenced by defects, thus we also tested the interaction of water with oxygen vacancies ($V_O$), see Fig. 5. In previous work we found that typically surface $V_O$s are not present on as-cleaved strontium ruthenate surfaces [34]. They can be created purposely by irradiating with electrons, however, as we have done in Fig. 5a. At low temperature water does not interact with the $V_O$s (Fig. 5b), but when a water-exposed surface is slightly annealed (Fig. 5c), bright species are observed. Water dissociating at O vacancies results in the vacancy being filled and two $O_{surf}H$ species. These are clearly distinct from the hydroxyl pairs (compare the dimer in Fig. 5d). They also form 1D chains, albeit with a large separation between them that points towards a repulsive interaction between isolated $O_{surf}H$.

**Discussion**

These results show that the theoretical concepts derived for rocksalt-type oxides can also be applied to AO-terminated perovskite surfaces. The confirmation of the theoretically-predicted, constricted motion of the $(OH)_{ads}$ fragment raises confidence in computational studies of water adsorption. Moreover, it allows distinguishing the state of single, isolated water molecules, even at very low coverage where they are truly isolated. As is expected for a rocksalt-type surface with larger lattice constant, the monomer dissociates [28]. The fact that the O octahedra on a perovskite surface are less rigid as compared to a rocksalt structure contributes to the propensity for dissociation. Probably this results in strong hydroxylation of SrO-terminated $SrTiO_3$(001) surfaces, as has been concluded from friction force measurements [41]. Recent near-ambient pressure XPS experiments report a rather complex O*1s* feature upon exposure to a humid atmosphere [42]. The XPS spectra in Fig. 4 and in the Supplement, taken on an as-cleaved surface and backed up by atomically-resolved STM results, provide an unequivocal interpretation of the various peaks.

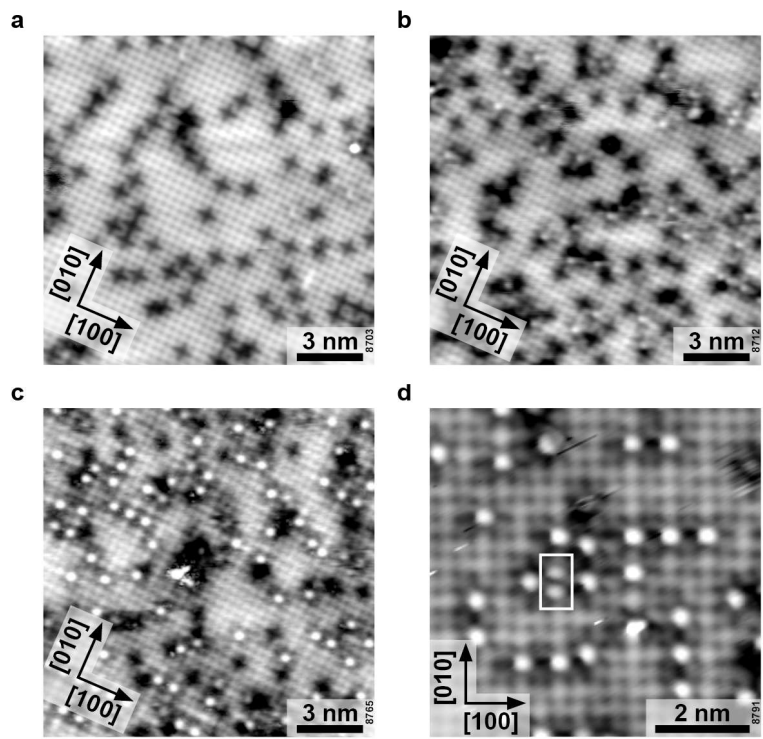

**Figure 5. Interaction of water with O vacancies.** a) $Sr_3Ru_2O_7$ after irradiation with 1000 eV electrons at 105 K (current ~ 1.5 µA). Vacancies appear as small, dark crosses. b) After exposure to 0.02 Langmuir of water at 105 K. c) As in (b) but after annealing to 200 K for 30 minutes. d) As in (c) but after annealing to 300 K for 30 minutes. Small-area image (image rotated 26.5° counterclockwise to scan direction). One dimer is marked with a box. Scanning conditions: $I_{tunnel}$ = 0.15 nA, $T_{sample}$ = 78 K, (a) $V_{sample}$ = +0.1 V, (b) $V_{sample}$ = +0.2 V, (c) and (d) $V_{sample}$ = +0.8 V.

While the structure and dynamics of the monomer is in line with the expectations for a SrO(001) surface, the configuration of the 'hydroxyl pair dimer' and the 1D chains is influenced by the $RuO_2$ layer underneath. The observed aggregation into dimers and 1D chains implies that the $(OH)_{ads}$ + H fragments associate easily so that the water diffuses as a whole unit. At this point it is unclear whether this diffusion is assisted by other water molecules as has been observed in several other cases [43,44]. While 1D chains have been observed on a CaO(001) surface [31], these were interpreted as aggregates of dissociated and non-dissociated water molecules interacting with each other. In case of SrO-terminated strontium ruthenates the hydroxyl pairs preferably sit at neighbouring sites that allow a tilt and rotation of the oxygen octahedra that is most accommodating to the adsorbates. This allows a continued, strong interaction between the fragments when the surface is exposed to more water: the chains of hydroxyl pairs grow, eventually forming a percolating network. Apparently the area right next to a 1D chain is not capable of dissociating the water; at higher exposures molecular water adsorbs in the gaps between the chains. This molecular water is adsorbed more weakly and desorbs below room temperature. This behaviour is different from the formation mechanism of the mixed dissocated/molecular water layer on MgO, where isolated water adsorbs molecularly [28,30], but water-water interaction is likely responsible for autocatalytic dissociation [29,32,33].

In summary, we have given a detailed picture of the first layer of water in direct contact with pristine and defect-free SrO-terminated perovskite surfaces. The results, taking under the most ideal and well-controlled conditions clearly show how $H_2O$ adsorption occurs from the monomer to the full layer. Dissociation is facile at regular lattice sites of SrO-terminated perovskites. The resulting OH and H fragments do not separate, however, but stay connected through an H bond between them, leading to an on-site diffusion process that is directly observed with STM. While the structural flexibility of the perovskite lattice initially facilitates water dissociation and hydroxylation, this only works up to a point. The lattice distortions ensuing from the adsorption limits this capacity, resulting in a mixed layer at higher coverages. The modification of the structural parameters, which is so successful for tuning the electrical and magnetic properties of perovskites, should thus also be instrumental in adjusting surface chemistry and reactivity of these promising materials.

**Materials and Methods**

**Experiment**

The experiments were carried out in a two-chamber UHV-system with base pressures of $2\times10^{-11}$ and $6\times10^{-12}$ mbar in the preparation chamber and the STM chamber, respectively. A low-temperature STM (commercial Omicron LT-STM) was operated at 78 K in constant-current mode using an electro-chemically etched W-tip. The bias voltage was applied to the sample; positive or negative bias voltages result in STM images of the unoccupied or occupied states, respectively. High-quality strontium ruthenate single crystals were grown by the floating zone technique using a two mirror image furnace. A detailed description of the growth procedure is found in ref. [45]. The samples were fixed on Omicron sample plates with conducting silver epoxy glue (EPO-TEK H21D), and a metal stud was glued on top with another epoxy adhesive (EPO-TEK H77). The crystals were cleaved by removing the metal stub with a wobble stick. Cleaving was performed in the analysis chamber at 100 K and 300 K with no apparent difference in sample quality [34]. Deionised water was further cleaned by several freeze-pump-thaw cycles and was dosed in the preparation chamber while keeping the sample at 105 K unless otherwise noted. Electron bombardment was performed by a well-outgassed electron gun in the preparation chamber with the sample held at 105 K. XPS was performed in a separate UHV chamber with a base pressure of $1\times10^{-10}$ mbar using monochromatized Al Kα X-rays and a SPECS PHOIBOS 150 electron analyser at normal emission with a pass energy of 16 eV.

**DFT**

The DFT calculations were performed with the Vienna Ab-initio Simulations Package (VASP) [46,47] in the PAW framework [48], using the optB86 van der Waals (vdW-DF) exchange-correlation functional [49]. The surface was modelled by a (4×4) surface cell (with respect to the unrotated octahedra) of a $Sr_3Ru_2O_7$ double layer terminating at the cleavage plane. The uppermost three layers were fully relaxed performing the Brillouin zone integration on a 3×3×1 Monkhorst-Pack k-point mesh. The transition state was determined with the dimer method [50], and subsequently verified by an explicit mapping of the reaction pathway.

**Acknowledgements**

This work has been supported by the ERC Advanced Grant 'OxideSurfaces' and by the Austrian Science Fund (FWF, Project F45). The Tulane team (DF, JP, and ZM) acknowledge support by the NSF under grant DMR-1205469. The Vienna Scientific Cluster is gratefully acknowledged for providing computing time. The authors thank Bilge Yildiz for useful discussions.

**Author contributions**

DH, BS, and MS performed the STM experiments and data analysis. DH, JPa, and GSP performed the XPS measurements. WMS, FM, and JR performed the DFT calculations. DF, JPe, and ZM grew the sample. UD directed and supervised the project. BS, DH, FM, MS, GSP, and UD wrote the manuscript.

**Competing financial interests**

The authors declare no competing financial interests.